\documentclass[aps,showpacs,twocolumn]{revtex4-1}
\usepackage{amssymb}
\usepackage{amsmath}
\usepackage{graphicx}
\usepackage{epsfig}
\usepackage{color}

\begin{document}

\title{Flat band and $\eta $-pairing states in one-dimensional Moir\'{e}
Hubbard model}
\author{R. Wang, and Z. Song}
\email{songtc@nankai.edu.cn}
\affiliation{School of Physics, Nankai University, Tianjin 300071, China}
\begin{abstract}
A moir\'{e} system is formed when two periodic structures have a slightly
mismatched period, resulting in unusual strongly correlated states in the
presence of particle-particle interactions. The periodic structures can
arise from the intrinsic crystalline order and periodic external field. We
investigate a one-dimensional Hubbard models with periodic on-site potential
of period $n_{0}$, which is commensurate to the lattice constant. For large $%
n_{0}$, exact solution demonstrates that there is a midgap flat band with
zero energy in the absence of Hubbard interaction. Each moir\'{e} unit cell
contributes two zero energy levels to the flat band. In the presence of
Hubbard interaction, the midgap physics is demonstrated to be well described
by a uniform Hubbard chain, in which the effective hopping and on-site
interaction strength, can be controlled by the amplitude and period of the
external field. Numerical simulations are performed to demonstrate the
correlated behaviors in the finite-sized moir\'{e} Hubbard system, including
the existence of $\eta $-pairing state, and bound pair oscillation. This
finding provides a method to enhance the correlated effect by a spatially
periodic external field.
\end{abstract}

\maketitle

\section{Introduction}

The theoretical and experimental studies on twisted bilayer graphene (TBG)
indicate that the interplay of lattice geometry and many-body interactions
can induce exotic quantum states \cite{RB,CY1,CY2,CY3,KK,YM,XL}, which
include superconducting and correlated insulating behavior. In general,
these phenomena are known to arise from the formation of flat band with low
energy and narrow bandwidth, which greatly enhances the electronic
interaction effect. However, it is a little tough to observe these
interesting correlation-induced phenomena because accurate magic angle in
TBG brings constraints on device fabrication. Therefore, it is natural to
look for a one-dimensional version of TBG, providing easier access to the
flat band by alternative approaches. A starting point is the moir\'{e}
patterns, which emerge due to the superposition of two periodic structures,
with either slightly different periods or different orientations, such
patterns have been realized in materials \cite%
{M.Yan,L.A.,C.R.,B.Hunt,C.Woods,YT,Bloch}.

In principle, a one-dimensional moir\'{e} system can be deliberately
engineered as a super-long-period system when the periods of two ingredients
are slightly different but commensurate \cite{YXM,WR1,WR2}. It is
essentially a one-dimensional periodic system with a large unit cell, which
may result in weak inter-unit-cell coupling, or nearly flat band.
Theoretically, two periodic structures can arise from the intrinsic
crystalline order and periodic external field. In condensed matter,
particles are usually confined by a periodic potential in synthetic and
natural materials. Another periodic structure can come from another external
field. Comparing to magic angle in TBG, the amplitude and period of a
periodic external field may be easy to control in experiment.

In this paper, we investigate a moir\'{e} Hubbard model, which is a uniform
Hubbard chain subjected to a field by cosinusoidal modulation with period
commensurate to the lattice constant. In the absence of the Hubbard
interaction, a midgap flat band is induced by the periodic field when the
field amplitude or the supperlattice unit cell is large enough. We determine
that the corresponding eigen wave functions are Gaussian-like with their
centers located at each zero potential point. When the field strength or
unit cell decreases, the flat band become narrow band, which can be well
characterized by tight-binding approximation, the Gaussian-like wave
functions playing the role of Wannier functions. In the presence of the
Hubbard interaction, the moir\'{e} pattern physics is demonstrated to be
well described by a uniform Hubbard chain with effective hopping ($J_{\text{%
eff}}$) and on-site interaction strength ($U_{\text{eff}}$). Importantly,
the ratio $U_{\text{eff}}/J_{\text{eff}}$ can be controlled by the amplitude
and period of the external field, to achieve the goal of enhancing the
electronic interaction effect. Numerical simulations are performed to
demonstrate the correlated behaviors of an ordinary Hubbard model in the
finite-sized moir\'{e} Hubbard system. The first one is the existence of $%
\eta $-pairing state, which is believed as a possible mechanism of
superconductivity based on the concept off-diagonal long-range order
(ODLRO). Another is the bound pair oscillation, demonstrating a correlated
dynamics. This finding provides a theoretical model as one-dimensional
analogue of twisted bilayer graphene. The advantage of this method is that
the correlated effect can be enhanced by the external field. We expect our
results benefit to experimental research.

This paper is organized as follows. In Section \ref{Model Hamiltonian}, we
present a one-dimensional Hubbard model with a spatially modulated on-site
potential. In Section \ref{Midgap Hubbard model}, we discuss midgap physics
in the model. In Section \ref{Bound-pair dynamics}, we propose a dynamic
scheme to experimentally demonstrate that the moir\'{e} Hubbard model
supports midgap bound-pair states. Finally, we provide a summary in Section %
\ref{Summary}.

\section{Model Hamiltonian}

\label{Model Hamiltonian} 
\begin{figure}[tbp]
\centering
\includegraphics[bb=111 268 491 491, width=0.5\textwidth, clip]{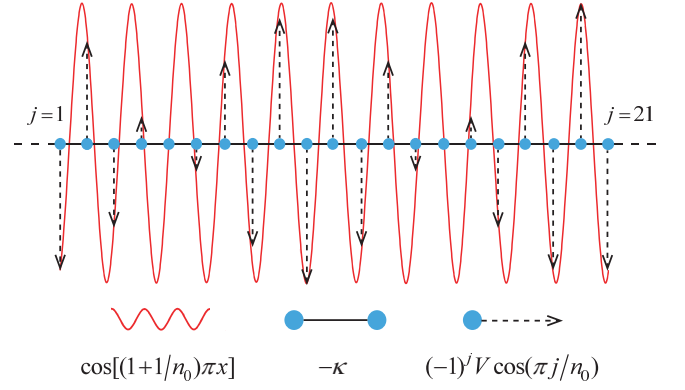}
\caption{Schematic illustration of a one-dimensional moir\'{e} system, which
consists two periodic ingredients with different but commensurate periods.
The lattice (array of solid blue circles) constant is $1$, while the
external field (red line) is cosinusoidal function $\cos [(1+1/n_{0})\protect%
\pi x]$ with period $2/(1+1/n_{0})$. The whole system is still periodic with
a long period $2n_{0}$ for large $n_{0}$. The dotted arrow denotes the
on-site potential distribution on the lattice, which is spatially modulated
and staggered as $(-1)^{j}\cos (\protect\pi j/n_{0})$ in Eq. (\protect\ref%
{H2}), where $j=\mathrm{INT}(x)$ is integer portion of the coordinate $x$.
Here $-\protect\kappa $ is nearest-neighbour hopping strength and $n_{0}=10$%
. }
\label{fig1}
\end{figure}

We consider a one-dimensional Hubbard model with a
spatially modulated on-site potential%
\begin{equation}
H=H_{0}+U\sum_{j}n_{j,\uparrow }n_{j,\downarrow }  \label{H1}
\end{equation}%
with%
\begin{eqnarray}
H_{0} &=&\sum_{j,\sigma =\uparrow ,\downarrow }[-\kappa c_{j,\sigma
}^{\dagger }c_{j+1,\sigma }+\mathrm{H.c.}+(-1)^{j}V  \notag \\
&&\times \cos (\frac{\pi j}{n_{0}})n_{j,\sigma }],  \label{H2}
\end{eqnarray}%
where $c_{j,\sigma }$ ($c_{j,\sigma }^{\dagger }$) is the annihilation
(creation) operator for an electron at site $j$ with spin $\sigma $ and $%
n_{j,\sigma }=c_{j,\sigma }^{\dagger }c_{j,\sigma }$. We schematically
illustrate that such a potential can be realized by a continuous potential
of trigonometric function (see Fig. \ref{fig1}). $H$ is known as a
one-dimensional standard Hubbard model with the uniform case ($V=0$), which
has been studied in many perspectives \cite{HT,VEK}. However, this
work investigates on the effect of nonzero $V$ on the property of Hubbard
model. Parameters $\kappa $ and $U$ play the role of kinetic and interaction
energy scale, respectively. The on-site staggered potential is commensurate
for integer $n_{0}$, which determines the Mori\'{e} period $2n_{0}$. We
consider a ring lattice of length $N=2Mn_{0}$ and take $c_{N+1,\sigma
}\equiv c_{1,\sigma }$ to impose a periodic boundary condition. The main
results of this work are independent of the boundary conditions.

We focus on the interaction-free case with zero $U$ and rewrite $H_{0}$ as
the form of%
\begin{eqnarray}
H_{0} &=&\sum_{m=1}^{M}[\sum_{l=1}^{2n_{0}-1}(-\kappa )(c_{m,l}^{\dagger
}c_{m,l+1}+c_{m,2n_{0}}^{\dagger }c_{m+1,1}+\text{\textrm{H.c.}})  \notag \\
&&+\sum_{l=1}^{2n_{0}}(-1)^{l}V\cos (\frac{\pi l}{n_{0}})n_{m,l}],
\label{H1_1}
\end{eqnarray}%
where we neglect the index of $\sigma $, with the mapping $%
c_{2(m-1)n_{0}+l,\sigma }=c_{m,l}$ and $n_{m,l}=c_{m,l}^{\dagger }c_{m,l}$.
Here index $m$ denotes the unit cell, while $l$ denotes the position of the
site in the $m$-th unit cell. We impose the periodic boundary condition $%
c_{M+1,1}=c_{1,1}$ and perform the Fourier transformations%
\begin{equation}
c_{m,l}=\frac{1}{\sqrt{M}}\sum_{k}e^{ikm}c_{k,l},
\end{equation}%
with $k=2n\pi /M$, $n=0,1,...,M-1$. Therefore, the transformation block
diagonalizes the Hamiltonian with translational symmetry, that is, $H_{0}$
can be written as block-diagonal form of%
\begin{equation}
H_{0}=\sum_{k}H_{0k}=\sum_{k}\psi _{k}^{\dagger }h_{0k}\psi _{k},
\label{H0k}
\end{equation}%
satisfying $[H_{0k},H_{0k^{\prime }}]=0$, in which the operator vector is%
\begin{equation}
\psi _{k}=(c_{k,1},c_{k,2},...,c_{k,2n_{0}})^{T},
\end{equation}%
and $h_{0k}$ is a $2n_{0}\times 2n_{0}$ matrix
\begin{figure*}[tbp]
\centering
\includegraphics[bb=75 22 2142 1600, width=1.0\textwidth, clip]{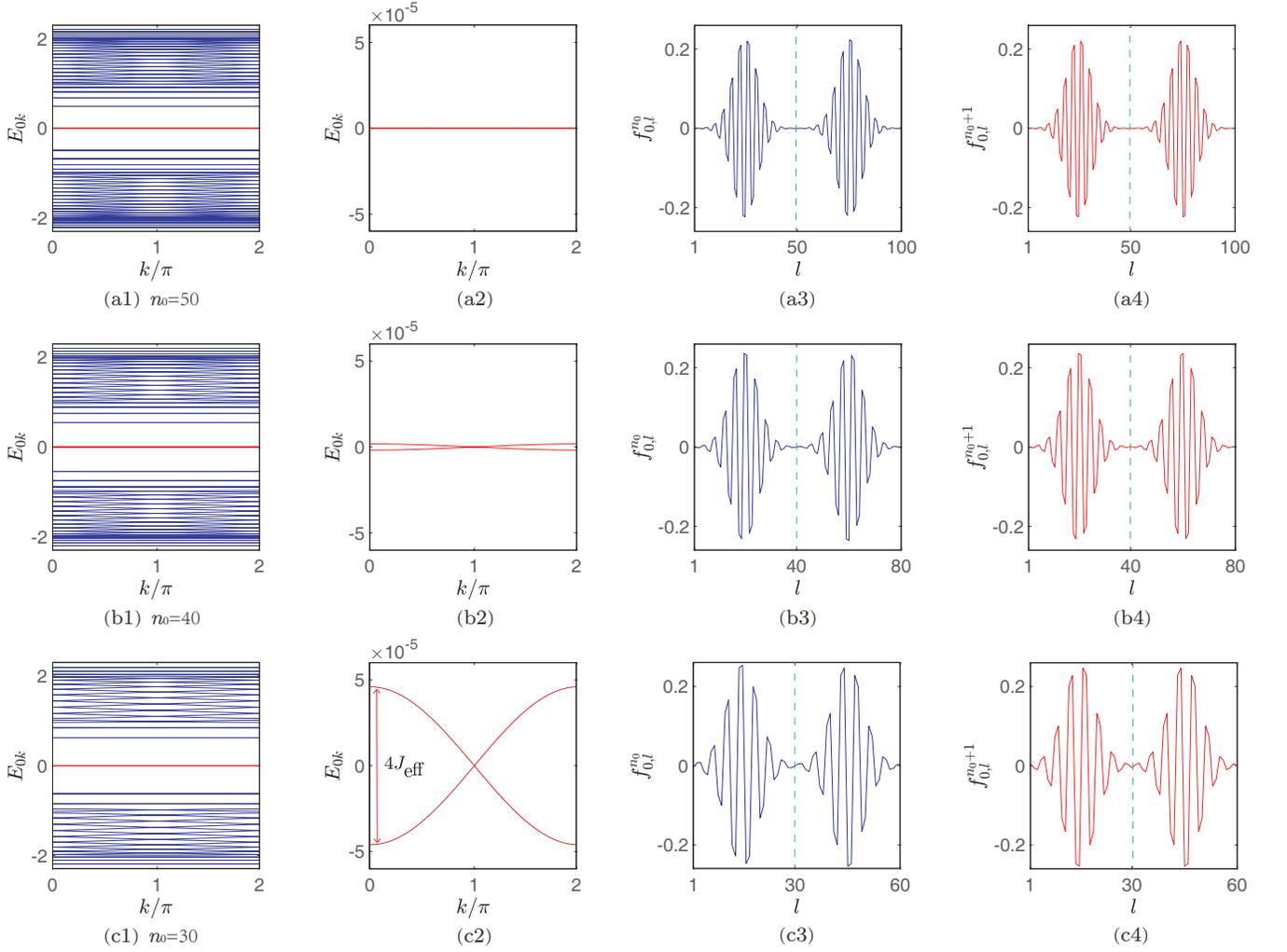}
\caption{Plots of energy spectra (a1-c1) and selected midgap wave functions
(a3-c3) and (a4-c4), obtained from exact diagonalization of the matrix in
Eq. (\protect\ref{h0k}) for fixed $V$ and three typical $n_{0}=50$, $40$,
and $30$, respectively. (a2-c2) is zoom of (a1-c1), selected typical energy
levels around zero energy. We only plot the wave functions in a single unit
cell with $k=0$. We determine that the zero energy states are antisymmetric
(blue) and symmetric (red) about the point $l=n_{0}$ (green dotted line),
which accords to the fact that $H_{0}^{\text{eff}}$ in Eq. (\protect\ref%
{H0eff}) has reflection symmetry. As $n_{0}$ decreases, the width of the
wavepackets increases and the overlap between two neighbour ones appears.
Meanwhile, the midgap flat band becomes midgap narrow band. The energy unit
is $\protect\kappa $. The system parameters are $\protect\kappa =V=1$, and $%
M=100$. }
\label{fig2}
\end{figure*}
\begin{equation}
h_{0k}=\left(
\begin{array}{ccccccc}
A_{1} & -\kappa & 0 & \cdots &  &  & -\kappa e^{-ik} \\
-\kappa & A_{2} & -\kappa & 0 & \cdots &  &  \\
0 & -\kappa & A_{3} & -\kappa & 0 & \cdots &  \\
\vdots & 0 & -\kappa & \ddots &  &  &  \\
& \vdots & 0 &  &  &  &  \\
&  & \vdots &  &  & \ddots & -\kappa \\
-\kappa e^{ik} &  &  &  &  & -\kappa & A_{2n_{0}}%
\end{array}%
\right)  \label{h0k}
\end{equation}%
with $A_{l}=(-1)^{l}V\cos (\pi l/n_{0})$. Matrix $h_{0k}$ corresponds to a
uniformly hopping ring with non-uniform on-site potential and a threading
flux $k$ through it. The flux can be neglected in large $n_{0}$ limit. The
system then has approximate reflection symmetry about the site $n_{0}$ due
to the fact $A_{2n_{0}-l}=A_{l}$. The eigenvector of $h_{0k}$ has the form of%
\begin{equation}
\varphi _{0}(k,n)=\sum_{l}^{2n_{0}}f_{k,l}^{n}c_{k,l}^{\dagger },
\end{equation}%
where $n=1,2,...,2n_{0}$, denotes the energy level index $E_{0k}(n)$. Exact
solution of $\left\{ f_{k,l}^{n}\right\} $ is tough to be obtained. However,
there are $M$ pairs of zero potential points at $(m,l)=(m,n_{0}/2)$ and $%
(m,3n_{0}/2)$ (or $j=2n_{0}(m-1)+n_{0}/2$ and $2n_{0}(m-1)+3n_{0}/2$ for the
original coordinate) in the limit case with $n_{0}\gg 1$ and nonzero $V$. In
the following, we will demonstrate that there are $M$ pairs of zero energy
eigenvectors.

As an approximation, we take a linearization of the Hamiltonian as effective
Hamiltonian by Taylor expansion of the potential around each zero potential
points. Defining $\Lambda $ as the range of validity for this linearization
approximation, we write the linearized effective Hamiltonian as the form of%
\begin{equation}
H_{0}^{\text{eff}}=\sum_{k}(H_{0k}^{+}+H_{0k}^{-}),  \label{H0eff}
\end{equation}%
where%
\begin{eqnarray}
H_{0k}^{\pm } &=&\sum_{l=n_{0}\pm n_{0}/2-\Lambda }^{n_{0}\pm
n_{0}/2+\Lambda }[-\kappa c_{k,l}^{\dagger }c_{k,l+1}+\text{\textrm{H.c.}}%
\mp (-1)^{l}V  \notag \\
&&\times (n_{0}\pm n_{0}/2-\frac{\pi l}{n_{0}})n_{k,l}].
\end{eqnarray}%
We determine that $H_{0}^{\text{eff}}$ has reflection symmetry about the
point $l=n_{0}$, as mentioned of the approximate symmetry of $h_{0k}$.
Introducing a set of operators defined by%
\begin{eqnarray}
\phi _{\pm }^{\dag }(k) &=&(\frac{V}{2\kappa n_{0}})^{1/4}\sum_{l=n_{0}\pm
n_{0}/2-\Lambda }^{n_{0}\pm n_{0}/2+\Lambda }\sin [(1+2l)\pi /4]  \notag \\
&&\times e^{-\frac{V\pi }{4\kappa n_{0}}[l-(n_{0}\pm
n_{0}/2)]^{2}}c_{k,l}^{\dagger },  \label{phi_k}
\end{eqnarray}%
a straightforward derivation demonstrates%
\begin{equation}
\lbrack H_{0}^{\text{eff}},\phi _{\pm }^{\dag }(k)]\approx 0.
\end{equation}%
It approximately indicates that a set of states $\phi _{\pm }^{\dag
}(k)\left\vert 0\right\rangle $ are eigenstates of $H_{0}^{\text{eff}}$ with
zero energy. Similarly, a set of local Gaussian wavepackets $\phi _{\pm
}^{\dag }(m)\left\vert 0\right\rangle $ with%
\begin{eqnarray}
\phi _{\pm }^{\dag }(m) &=&(\frac{V}{2\kappa n_{0}})^{1/4}\sum_{l=n_{0}\pm
n_{0}/2-\Lambda }^{n_{0}\pm n_{0}/2+\Lambda }\sin [(1+2l)\pi /4]  \notag \\
&&\times e^{-\frac{V\pi }{4\kappa n_{0}}[l-(n_{0}\pm
n_{0}/2)]^{2}}e^{ikm}c_{m,l}^{\dagger }  \label{GS_WF}
\end{eqnarray}%
are eigenstates of $H_{0}^{\text{eff}}$ with zero energy, where $\left\vert
0\right\rangle $ is the vacuum state for the creation fermion operator $%
c_{m,l}^{\dagger }$. It is easy to check that there are totally $2M$ zero
energy eigenvectors.

\begin{figure}[tbp]
\centering
\includegraphics[bb=16 44 580 1545, width=0.4\textwidth, clip]{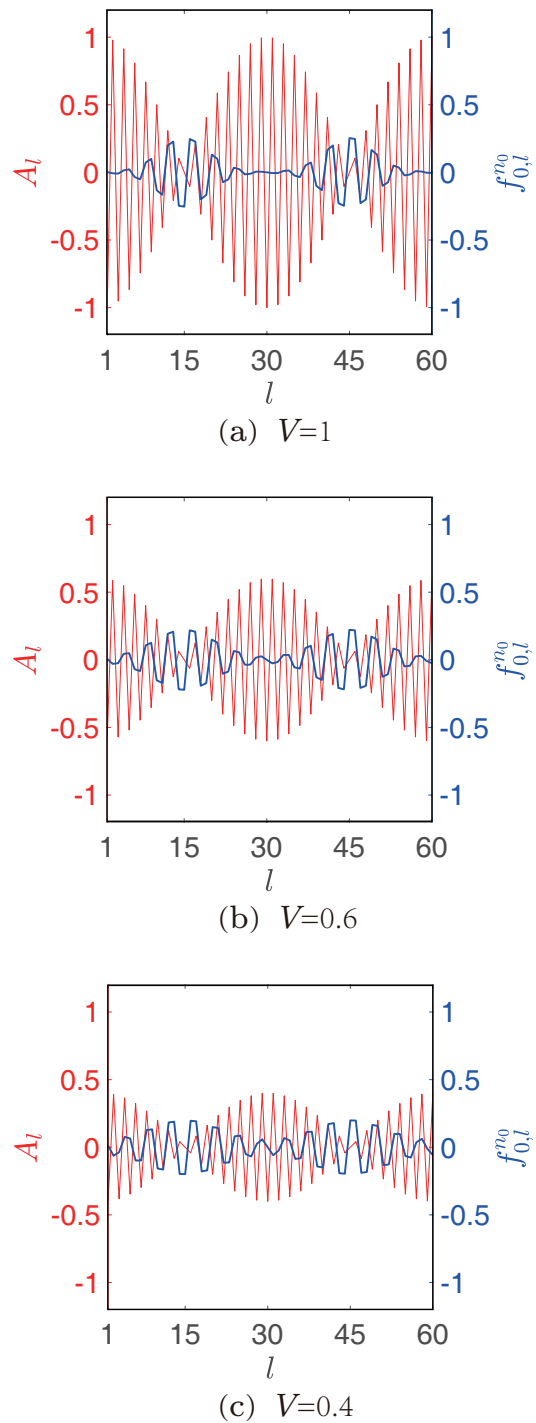}
\caption{Plots of the profiles of wave functions (blue) in the midgap and
on-site potential distribution $A_{l}$ (red). Three representative values of
$V$ are selected and the corresponding wave functions $f_{k,l}^{n}$ with $%
k=0 $ are obtained by exact diagonalization of matrix in Eq. (\protect\ref%
{h0k}). As expected, we see that the wave functions are Gaussian-like with
the center being localized around zero potential points. The width of
wavepackets increases and the overlap between two neighbour ones appears as $%
V$ decreases. The system parameters are $\protect\kappa =1$, and $n_{0}=30$.}
\label{fig3}
\end{figure}

Eigenvector $\varphi _{0}(k,n)$ and spectrum $E_{0k}(n)$ can be obtained
numerically by exact diagonalization of the matrix $h_{0k}$ in the case of
finite $n_{0}$ limit. We plot the profiles of the wave function and the
spectrum for three representative cases (see Fig. \ref{fig2}). We determine
that for sufficient large $n_{0}$, the zero energy states are separated
Gaussian wavepackets, which accords to our above analysis. As $n_{0}$
decreases, the width of the wavepackets increases and the overlap between
two neighbour ones appears. Moreover, the midgap flat band becomes midgap
narrow band. Similar things happen as fix $n_{0}$ and vary $V$: there is a
midgap flat band in large $V$ limit with $V\gg \kappa $ because every zero
potential sites decouple from the neighbours. Extreme localized states (site
state) form at each zero potential points. As $V$ decreases, the site state
becomes wavepacket and as its width increases, the overlap between two
neighbour ones appears. Each zero potential points become the centers of
wavepackets. Moreover, the midgap flat band becomes midgap narrow band. This
process is demonstrated by numerical simulation in Fig. \ref{fig3}.

It is clear that this moir\'{e} system can be regarded as a superlattice: $%
\phi _{\pm }(m)$ [$\phi _{\pm }^{\dag }(m)$] annihilates (creates) a
localized Wannier state at superlattice site $m$. The midgap band is
depicted by tight-binding Hamiltonian%
\begin{eqnarray}
H_{0}^{\text{MG}} &=&J_{\text{eff}}\sum_{m}[\phi _{+}^{\dag }(m)\phi _{-}(m)
\notag \\
&&+\phi _{+}^{\dag }(m)\phi _{-}(m+1)+\mathrm{H.c.}].
\end{eqnarray}%
The hopping constant $J_{\text{eff}}$ can be estimated from numerical result
for the width ($4J_{\text{eff}}$) of midgap narrow band as
indicated in Fig. \ref{fig2}(c2). We would like to point out that, unlike
the usual single lowest band approximation, $J_{\text{eff}}$ is always
positive, since $H_{0}^{\text{MG}}$ describes a midgap band, does not
violate the node theorem for the eigen wave function in one-dimensional
system.
\begin{figure*}[tbp]
\centering
\includegraphics[bb=40 35 1700 1039, width=1.0\textwidth, clip]{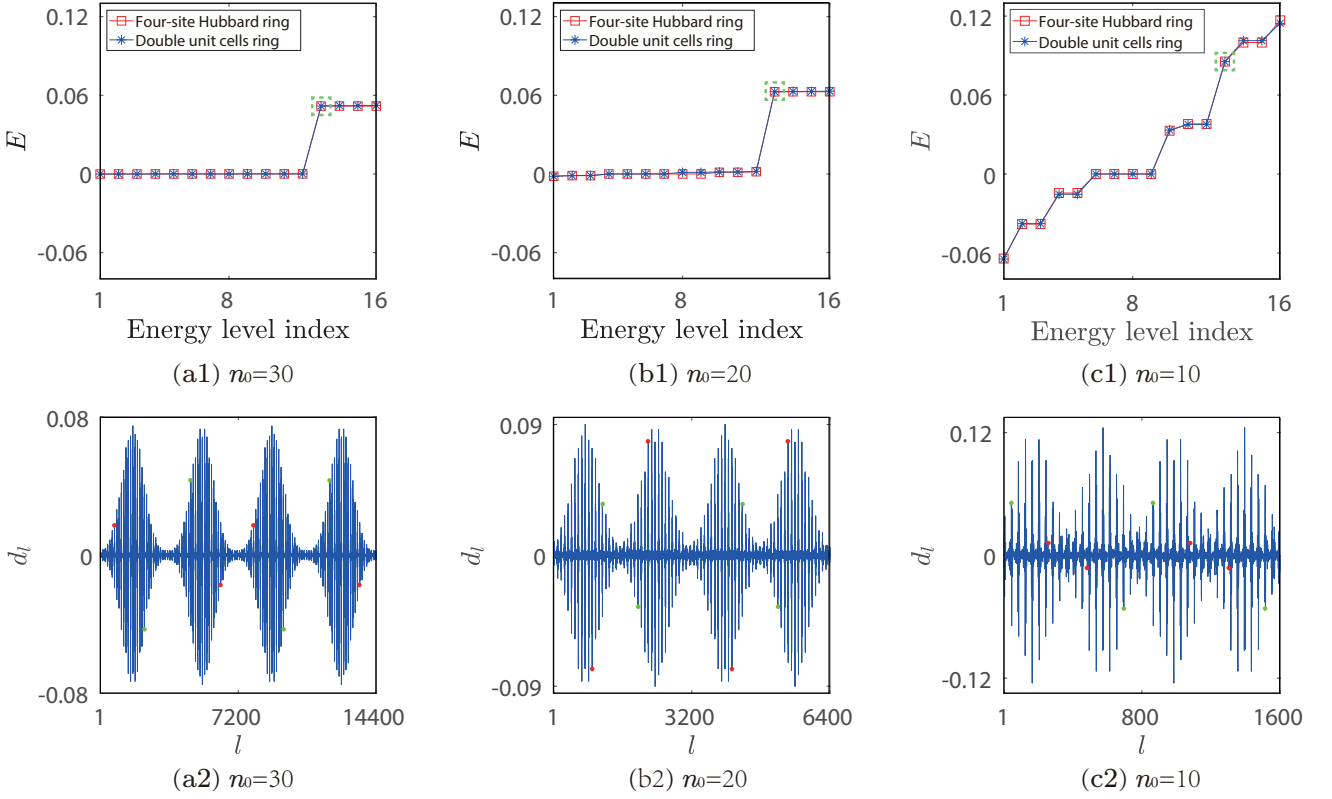}
\caption{Comparison of the midgap states of moir\'{e} Hubbard model and its
tight-binding description. (a1-c1) Plots of midgap energy levels (blue star)
obtained by exact diagonalization of the Hamiltonian in Eq. (\protect\ref%
{H_DUC}) for three typical $n_{0}$. Red squares stand for the energy levels
of a four-site Hubbard ring with fitting values of $J_{\text{eff}}$ and $U_{%
\text{eff}}$. We can find that for sufficient large $n_{0}$, the original
single flat band ($U=0$) separates into two flat bands (a1), the upper band
is pair band. As $n_{0}$ decreases, the width of the two bands increase
[(b1) and (c1)]. The pair-band is fitted by the numerical result from the
four-site Hubbard ring in Eq. (\protect\ref{H_MG}). The fitting parameters
of four-site Hubbard ring are $J_{\text{eff}}=2.30\times 10^{-5}$, $U_{\text{%
eff}}=5.18\times 10^{-2}$ ($5.11\times 10^{-2}$) in (a1), $J_{\text{eff}%
}=6.20\times 10^{-4}$, $U_{\text{eff}}=6.27\times 10^{-2}$ ($6.14\times
10^{-2}$) in (b1), and $J_{\text{eff}}=1.90\times 10^{-2}$, $U_{\text{eff}%
}=8.54\times 10^{-2}$ ($8.42\times 10^{-2}$) in (c1), respectively. Here the
values of $U_{\text{eff}}$ in each bracket are obtained from Eq. (\protect
\ref{U_eff}). (a2-c2) Plots of the eigen wave function $d_{l}$ of the lowest
energy level, i.e., $E=U_{\text{eff}}$ (green dashed box) at the upper band.
It is the $\protect\eta $-pairing state, which obeys the relation in Eq. (%
\protect\ref{symm}) as expected, indicating by two groups of red and green
solid circles. The results demonstrate that the tight-binding approximation
works quite well and there exist $\protect\eta $-pairing states in the
supperlattice near the Fermi level. The system parameters are $\protect%
\kappa =1$, $V=1$, and $U=0.5$.}
\label{fig4}
\end{figure*}

\section{Midgap Hubbard model}

\label{Midgap Hubbard model}

So far we have established the tight binding description of the midgap
physics for the moir\'{e} system. In the following section, we will consider
the case with nonzero $U$. $H_{0}^{\text{MG}}$ then can be modified by
adding the spin index, replacing $\phi _{\pm }(m)$ by $\phi _{\pm ,\sigma
}(m)$. The above analysis still holds for interaction-free terms ($%
H_{0,\sigma }^{\text{MG}}$). The midgap Hubbard model Hamiltonian reads%
\begin{eqnarray}
H^{\text{MG}} &=&\sum_{\sigma =\uparrow ,\downarrow }H_{0,\sigma }^{\text{MG}%
}+U_{\text{eff}}\sum_{m,\rho =\pm }\phi _{\rho ,\uparrow }^{\dag }(m)\phi
_{\rho ,\uparrow }(m)  \notag \\
&&\times \phi _{\rho ,\downarrow }^{\dag }(m)\phi _{\rho ,\downarrow }(m),
\label{H_MG}
\end{eqnarray}%
where $H_{0,\sigma }^{\text{MG}}$ is extended from $H_{0}^{\text{MG}}$,
which has the form of%
\begin{eqnarray}
H_{0,\sigma }^{\text{MG}} &=&J_{\text{eff}}\sum_{m}[\phi _{+,\sigma }^{\dag
}(m)\phi _{-,\sigma }(m)  \notag \\
&&+\phi _{+,\sigma }^{\dag }(m)\phi _{-,\sigma }(m+1)+\mathrm{H.c.}].
\end{eqnarray}%
The effective Hubbard interaction strength $U_{\text{eff}}$ is determined by%
\begin{eqnarray}
U_{\text{eff}} &=&U\left\langle 0\right\vert \phi _{\rho ,\downarrow
}(m)\phi _{\rho ,\uparrow }(m)[\sum_{l}(n_{m,l,\uparrow }n_{m,l,\downarrow
})]  \notag \\
&&\times \phi _{\rho ,\uparrow }^{\dag }(m)\phi _{\rho ,\downarrow }^{\dag
}(m)\left\vert 0\right\rangle ,  \label{U_eff}
\end{eqnarray}%
which can be estimated by numerical computation. We plot the profiles of the
two-particle eigenstates and the spectrum for three representative cases
(see Fig. \ref{fig4}). We determine that for sufficient large $n_{0}$, the
flat band separates into two flat bands, one is pair-band, another is
scattering band. The width of the two bands increases as $n_{0}$ decreases.

To verify the tight-binding approximation, one can fit the result from exact
diagonalization for original moir\'{e} Hamiltonian by that of $H^{\text{MG}}$%
. We consider a system with double moir\'{e} unit cells,
\begin{eqnarray}
H_{M2} &=&\sum_{j=1,\sigma =\uparrow ,\downarrow }^{4n_{0}}[-\kappa
c_{j,\sigma }^{\dagger }c_{j+1,\sigma }+\mathrm{H.c.}+(-1)^{j}V  \notag \\
&&\times \cos (\frac{\pi j}{n_{0}})n_{j,\sigma
}]+U\sum_{j=1}^{4n_{0}}n_{j,\uparrow }n_{j,\downarrow },  \label{H_DUC}
\end{eqnarray}%
which corresponds to a four-site Hubbard ring.

Besides the Eq. (\ref{U_eff}), $U_{\text{eff}}$ can further be obtained by
fitting the energy levels of $H_{M2}$ to that of $H^{\text{MG}}$. The
fitting results are given in Fig. \ref{fig4}(a1-c1), and the obtained $U_{%
\text{eff}}$ by fitting is compared with the value estimated from Eq. (\ref%
{U_eff}). We determine that the approximation works quite well. Furthermore,
the bound-pair levels [upper band in Fig. \ref{fig4}(a1-c1)] accords with
the analytical expression
\begin{equation}
\varepsilon _{n}=\sqrt{U_{\text{eff}}^{2}+16J_{\text{eff}}^{2}\cos ^{2}(%
\frac{n\pi }{M})},  \label{JL}
\end{equation}%
with $n=0,1,...,2M-1$, which is obtained in the previous works \cite{JL1,JL2}%
.

Particularly, when taking $n=M/2$, we have $\varepsilon _{M/2}=U_{\text{eff}%
} $, which is the lowest bound-pair energy level. The corresponding
eigenstate is%
\begin{equation}
|\Psi _{\eta }^{1}\rangle =\eta \left\vert 0\right\rangle ,
\end{equation}%
where $\eta $ is the so-called $\eta $-pairing operator defined as%
\begin{equation}
\eta =\frac{1}{\sqrt{2M}}\sum_{m=1,\rho =\pm }^{M}\rho \phi _{\rho ,\uparrow
}^{\dag }(m)\phi _{\rho ,\downarrow }^{\dag }(m).
\end{equation}%
According to the seminal work by Yang \cite{Yang1989}, we have
\begin{equation}
\lbrack H^{\text{MG}},\eta ]=U_{\text{eff}}\eta ,
\end{equation}%
which ensures that%
\begin{equation}
H^{\text{MG}}|\Psi _{\eta }^{n}\rangle =nU_{\text{eff}}|\Psi _{\eta
}^{n}\rangle ,
\end{equation}%
i.e., state%
\begin{equation}
|\Psi _{\eta }^{n}\rangle =\eta ^{n}\left\vert 0\right\rangle ,
\end{equation}%
is the eigenstates of $H^{\text{MG}}$ with energy $nU_{\text{eff}}$. The $%
\eta $-pairing state $|\Psi _{\eta }^{n}\rangle $ has been demonstrated to
have ODLRO if $n/M$ is finite ($0<n/M<1$) in the thermodynamic limit, $%
M\rightarrow \infty $.

In general, a complete set basis of the two-electron subspace $\left\{
|l\rangle \right\} $ is%
\begin{equation}
|l\rangle =c_{i,\uparrow }^{\dagger }c_{j,\downarrow }^{\dagger }|0\rangle ,
\end{equation}%
where $l$ is defined as $l=2Mn_{0}(i-1)+j$, with $i,j\in \lbrack 1,2Mn_{0}]$%
. Any two-electron states with opposite spins then can be expressed in the
form of%
\begin{equation}
|\Psi \rangle =\sum_{l}d_{l}|l\rangle .
\end{equation}%
For small size moir\'{e} Hubbard model $H_{M2}$ with a pair of electrons,
the corresponding $\eta $-pairing state is
\begin{equation}
|\Psi _{\eta }^{1}\rangle =\frac{1}{2}\sum_{\rho =\pm }\rho \lbrack \phi
_{\rho ,\uparrow }^{\dag }(1)\phi _{\rho ,\downarrow }^{\dag }(1)+\phi
_{\rho ,\uparrow }^{\dag }(2)\phi _{\rho ,\downarrow }^{\dag }(2)]\left\vert
0\right\rangle ,
\end{equation}%
which requires the eigen wave function with energy $U_{\text{eff}}$
satisfies the relation%
\begin{equation}
\begin{array}{cc}
\begin{array}{c}
d_{l}=-d_{4(Mn_{0})^{2}-l} \\
d_{l}=d_{l+2(Mn_{0})^{2}}%
\end{array}%
, & l\in \lbrack 1,2(Mn_{0})^{2}].%
\end{array}
\label{symm}
\end{equation}%
Numerical results for eigen wave function plotted in Fig. \ref{fig4}%
(a2)-(c2) accord with this analysis. These findings demonstrate that midgap
physics for the Hubbard moir\'{e} system can be well captured by a simple
Hubbard model. The advantage of this system are evident: the parameters $J_{%
\text{eff}}$ and $U_{\text{eff}}$ can be controlled by the external field,
similar to what happens in optical lattice system. Moreover, midgap Hubbard
model provides a possibility to realize one-dimensional superconducting
state based on the formation of $\eta $ pairs near the Fermi level.

\section{Bound-pair dynamics}

\label{Bound-pair dynamics}

The above analysis indicates that the moir\'{e} Hubbard model supports
midgap bound-pair states, which are protected from thermal fluctuation due
to energy gap. This may be allowed to observe the related dynamics in
experiment, similar to the Bloch oscillation \cite{GC}. To verify and
demonstrate such a prediction, we consider a simplest case which is
consisted of a single moir\'{e} unit cell and investigate the dynamics of
the system with specific parameters.

The corresponding tight-binding Hamiltonian reads
\begin{eqnarray}
H_{\text{1}}^{\text{MG}} &=&J_{\text{eff}}[\phi _{+}^{\dag }(1)\phi _{-}(1)+%
\mathrm{H.c.}]+U_{\text{eff}}  \notag \\
&&\times \sum_{\rho =\pm }\phi _{\rho ,\uparrow }^{\dag }(1)\phi _{\rho
,\uparrow }(1)\phi _{\rho ,\downarrow }^{\dag }(1)\phi _{\rho ,\downarrow
}(1),
\end{eqnarray}%
which is a two-site Hubbard model. In the two-particle invariant subspace,
spaned by two-particle basis
\begin{equation}
\left\{
\begin{array}{c}
\phi _{+,\uparrow }^{\dag }(1)\phi _{+,\downarrow }^{\dag }(1)|0\rangle \\
\phi _{-,\uparrow }^{\dag }(1)\phi _{-,\downarrow }^{\dag }(1)|0\rangle \\
\phi _{+,\uparrow }^{\dag }(1)\phi _{-,\downarrow }^{\dag }(1)|0\rangle \\
\phi _{-,\uparrow }^{\dag }(1)\phi _{+,\downarrow }^{\dag }(1)|0\rangle%
\end{array}%
\right. ,  \label{basis}
\end{equation}%
the matrix representation of $H_{\text{1}}^{\text{MG}}$ is
\begin{equation}
h_{1}=J_{\text{eff}}\left(
\begin{array}{cccc}
U_{\text{eff}}/J_{\text{eff}} & 0 & 1 & 1 \\
0 & U_{\text{eff}}/J_{\text{eff}} & 1 & 1 \\
1 & 1 & 0 & 0 \\
1 & 1 & 0 & 0%
\end{array}%
\right) ,
\end{equation}%
which has parity symmetry, i.e., $Ph_{1}P^{-1}=h_{1}$. The reflectional
operator $P$ is defined as $P\phi _{\rho ,\sigma }^{\dag }(1)P^{-1}=\phi _{%
\overline{\rho },\sigma }^{\dag }(1)$ with $\rho =-\overline{\rho }=+,-$ and
$\sigma =\uparrow ,\downarrow $. Its matrix representation is

\begin{equation}
P=\left(
\begin{array}{cccc}
0 & 1 & 0 & 0 \\
1 & 0 & 0 & 0 \\
0 & 0 & 0 & 1 \\
0 & 0 & 1 & 0%
\end{array}%
\right) ,
\end{equation}%
in the basis set of Eq. (\ref{basis}). Therefore, four eigenvectors have
even or odd parity, i.e.,
\begin{equation}
\left\{
\begin{array}{c}
P\left\vert \psi _{1}\right\rangle =\left\vert \psi _{1}\right\rangle \\
P\left\vert \psi _{2}\right\rangle =-\left\vert \psi _{2}\right\rangle \\
P\left\vert \psi _{3}\right\rangle =-\left\vert \psi _{3}\right\rangle \\
P\left\vert \psi _{4}\right\rangle =\left\vert \psi _{4}\right\rangle%
\end{array}%
\right. ,
\end{equation}%
\begin{figure}[tbp]
\centering
\includegraphics[bb=59 31 1121 1535, width=0.5\textwidth, clip]{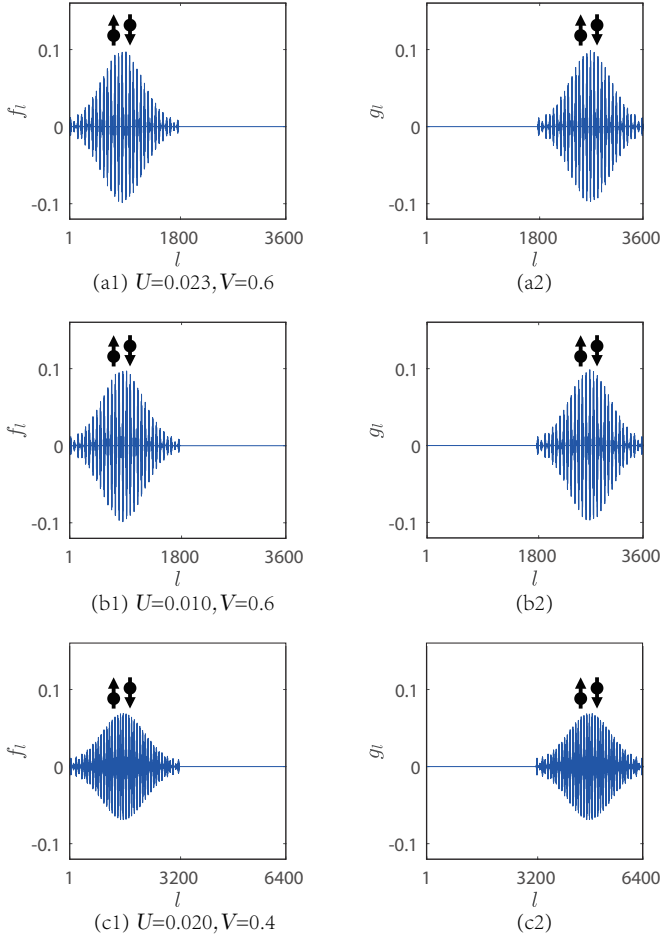}
\caption{Profiles of the initial and target states in Eq. (\protect\ref{fg}%
), the coefficients $f_{l}$ and $g_{l}$ with $\protect\kappa =\protect\kappa %
_{0}$, obtained by exact diagonalization. Plots of $f_{l}$ in (a1-c1) and $%
g_{l}$ in (a2-c2) for three typical situations. The system parameters are $%
n_{0}=30$, $U=0.023$, $V=0.6$ in (a1-a2), $n_{0}=30$, $U=0.010$, $V=0.6$ in
(b1-b2), $n0=40$, $U=0.020$, $V=0.4$ in (c1-c2), and $\protect\kappa =1$.}
\label{fig5}
\end{figure}
\begin{figure*}[tbp]
\centering
\includegraphics[bb=60 16 2085 1609, width=1.0\textwidth, clip]{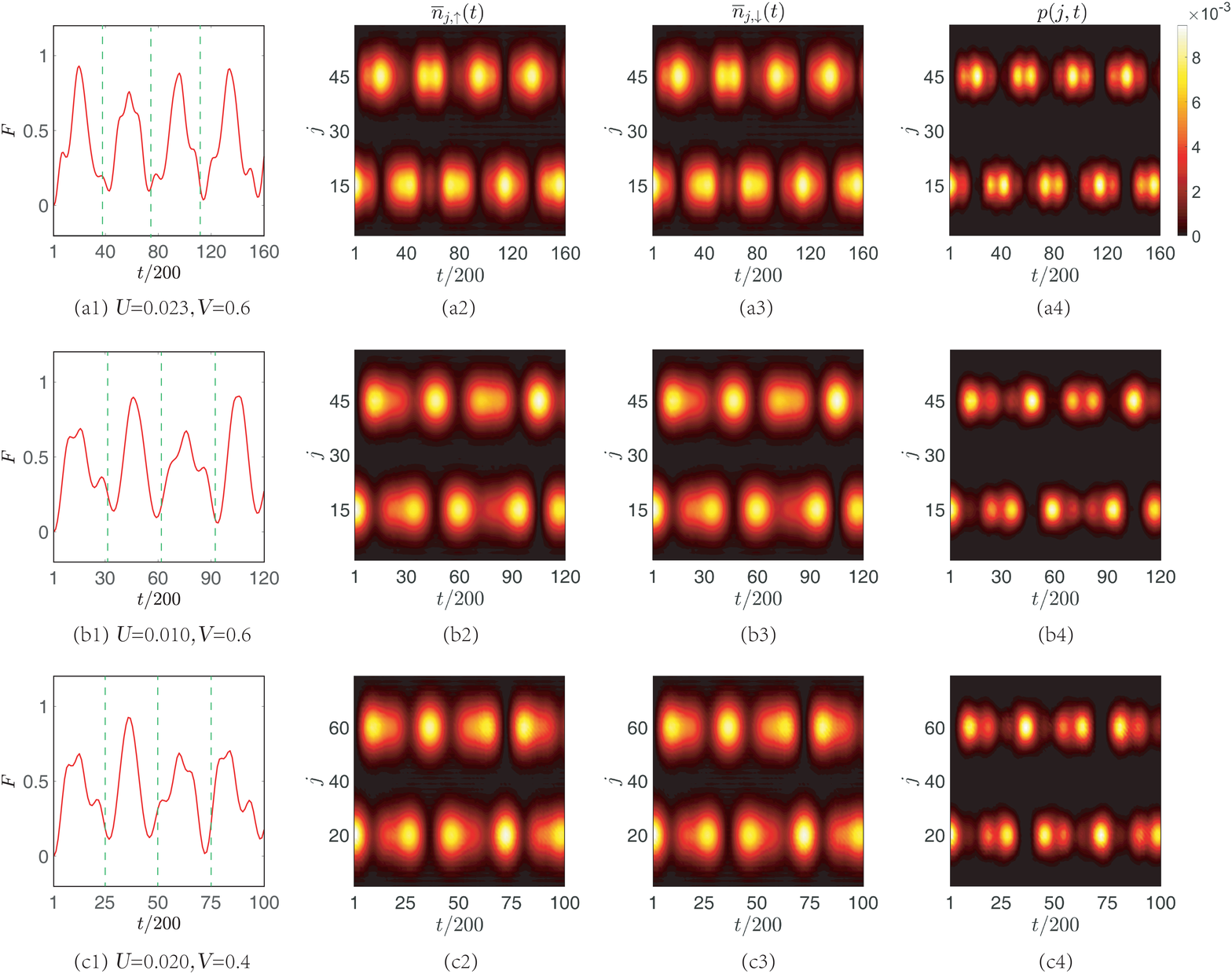}
\caption{Numerical results for the dynamical process of pair tunneling,
obtained by exact diagonalization. Four quantities, (a1-c1) fidelity ($F$),
(a2-c2) $\overline{n}_{j,\uparrow }(t)$, (a3-c3) $\overline{n}_{j,\downarrow
}(t)$, and (a4-c4) $p(j,t)$, defined in Eq. (\protect\ref{fidelity}) are
plotted for the evolved states in systems with four typical parameters. We
determine that $F$, $\overline{n}_{j,\uparrow }$, $\overline{n}%
_{j,\downarrow }$ and $p$ oscillate approximately with the period indicated
by green dotted lines in (a1-c1), obtained from analytical expression $T=2%
\protect\pi /\Delta $. Although the fidelity is not perfect, the pair
tunneling is evident. The results indicate that the tight-binding
approximation works quite well and a bound pair can be near perfectly
transferred between two unit cells of superlattice. The system parameters
are $n_{0}=30$, $\Delta =8.3\times 10^{-4}$, $U=0.023$, $V=0.6$ in (a1-a4), $%
n_{0}=30$, $\Delta =1.1\times 10^{-3}$, $U=0.010$, $V=0.6$ in (b1-b4), and $%
n0=40$, $\Delta =1.3\times 10^{-3}$, $U=0.020$, $V=0.4$ in (c1-c4). The time
unit is $\protect\kappa ^{-1}$ and we take $\protect\kappa =1$.}
\label{fig6}
\end{figure*}
where the eigenvalues and eigenvectors are
\begin{equation}
\begin{array}{c}
\varepsilon _{2}=U_{\text{eff}},\varepsilon _{3}=0, \\
\varepsilon _{1,4}=\frac{1}{2}(U_{\text{eff}}\pm \sqrt{U_{\text{eff}%
}^{2}+16J_{\text{eff}}^{2}}),%
\end{array}%
\end{equation}%
and%
\begin{equation}
\begin{array}{c}
\left\vert \psi _{2}\right\rangle =2^{-1/2}(%
\begin{array}{cccc}
-1 & 1 & 0 & 0%
\end{array}%
)^{T}, \\
\left\vert \psi _{3}\right\rangle =2^{-1/2}(%
\begin{array}{cccc}
0 & 0 & -1 & 1%
\end{array}%
)^{T}, \\
\left\vert \psi _{1,4}\right\rangle =(8+2|\varepsilon _{4,1}|^{2})^{-1/2}(%
\begin{array}{cccc}
-2 & -2 & \varepsilon _{4,1} & \varepsilon _{4,1}%
\end{array}%
)^{T},%
\end{array}%
\end{equation}%
respectively. This system allows a special dynamics if we take
\begin{equation}
U_{\text{eff}}=2\Delta /(2n+1)=4J_{\text{eff}}/\sqrt{(2n+1)(2n+3)},
\label{ratio}
\end{equation}%
at which we simply have%
\begin{equation}
\varepsilon _{1}=(\frac{2n+3}{2n+1})\Delta ,\varepsilon _{2}=\frac{2}{2n+1}%
\Delta ,\varepsilon _{3}=0,\varepsilon _{4}=-\Delta ,  \label{levels}
\end{equation}%
where $n=0,1,2,...$. Such a commensurate energy level structure, together
with the parities of the eigenvectors, allows the time evolution operator $%
\mathcal{U}(t)=e^{-ih_{1}t}$ to obey the connection to the parity operator

\begin{equation}
\mathcal{U}(t_{q})=\left( -P\right) ^{q}
\end{equation}%
at instant $t_{q}=q(2n+1)\pi /\Delta $. An arbitrary initial state can
evolve naturally to its parity counterpart or itself periodically. For
example, a Hubbard-pair initial state $\phi _{+,\uparrow }^{\dag }(1)\phi
_{+,\downarrow }^{\dag }(1)|0\rangle $ can evolve to another Hubbard-pair
state $\phi _{-,\uparrow }^{\dag }(1)\phi _{-,\downarrow }^{\dag
}(1)|0\rangle $, realizing pair tunneling.

We investigate the tight-binding approximation by numerical simulation of
the dynamics in the original system with Eq. (\ref{H1}). We consider the
Hamiltonian of a single moir\'{e} unit cell,

\begin{eqnarray}
H_{M1} &=&\sum_{j=1,\sigma =\uparrow ,\downarrow }^{2n_{0}}[-\kappa
c_{j,\sigma }^{\dagger }c_{j+1,\sigma }+\mathrm{H.c.}+(-1)^{j}V  \notag \\
&&\times \cos (\frac{\pi j}{n_{0}})n_{j,\sigma
}]+U\sum_{j=1}^{2n_{0}}n_{j,\uparrow }n_{j,\downarrow }  \notag \\
&&+\kappa _{0}\sum_{\sigma =\uparrow ,\downarrow }(c_{n_{0},\sigma
}^{\dagger }c_{n_{0}+1,\sigma }+\mathrm{H.c.}),  \label{H_SUC}
\end{eqnarray}%
where the $\kappa _{0}$ term controls the connection between to halves of
the chain: when taking $\kappa _{0}=\kappa $, we have $J_{\text{eff}}=0$.
The computation is performed in two-particle subspace, where two particles
have opposite spins. The subspace is spanned by a set of basis $%
\{c_{i,\uparrow }^{\dag }c_{j,\downarrow }^{\dag }|0\rangle
,i,j=1,...,2n_{0}\}$ of $\left( 2n_{0}\right) ^{2}$ dimensions.

We compute the initial state $|\psi _{\text{I}}\rangle $ and target state $%
|\psi _{\text{T}}\rangle $, by exact diagonalization method for the system
of Eq. (\ref{H_SUC}) with $\kappa _{0}=\kappa $. The initial state $|\psi _{%
\text{I}}\rangle $ and target state $|\psi _{\text{T}}\rangle $ has the form
\begin{equation}
\begin{array}{c}
|\psi _{\text{I}}\rangle =\sum_{l}f_{l}c_{i,\uparrow }^{\dag
}c_{j,\downarrow }^{\dag }|0\rangle , \\
|\psi _{\text{T}}\rangle =\sum_{l}g_{l}c_{i,\uparrow }^{\dag
}c_{j,\downarrow }^{\dag }|0\rangle ,%
\end{array}
\label{fg}
\end{equation}%
where $l=2n_{0}(i-1)+j$. Two states $|\psi _{\text{I}}\rangle $ and $|\psi _{%
\text{T}}\rangle $ are mutual parity counterpart due to the symmetry of $%
H_{M1}$. Taking $\kappa _{0}=0$, $|\psi _{\text{I}}\rangle $ can be set as a
initial state for time evolution and $|\psi _{\text{I}}\rangle $ changes to $%
|\psi _{\text{T}}\rangle $ by sufficient long evolution time. Proper
parameters ($U$ and $V$) are searched to achieve the commensurate energy
level structure as in Eq. (\ref{levels}). We plot the schematic illustration
of the initial state $|\psi _{\text{I}}\rangle $ and target state $|\psi _{%
\text{T}}\rangle $ for three typical conditions (see Fig. \ref{fig5}).
Obviously, the initial state $|\psi _{\text{I}}\rangle $ and target state $%
|\psi _{\text{T}}\rangle $ are mutual parity counterpart. The time evolution
is computed for $\kappa _{0}=0$ by the formula%
\begin{equation}
\left\vert \psi (t)\right\rangle =\exp (-iH_{1M}t)|\psi _{\text{I}}\rangle .
\end{equation}%
We define four quantities
\begin{equation}
\begin{array}{c}
F(t)=\left\vert \langle \psi _{\text{T}}\left\vert \psi (t)\right\rangle
\right\vert , \\
\overline{n}_{j,\uparrow }(t)=\left\langle \psi (t)\right\vert n_{j,\uparrow
}\left\vert \psi (t)\right\rangle , \\
\overline{n}_{j,\downarrow }(t)=\left\langle \psi (t)\right\vert
n_{j,\downarrow }\left\vert \psi (t)\right\rangle , \\
p(j,t)=\left\langle \psi (t)\right\vert n_{j,\uparrow }n_{j,\downarrow
}\left\vert \psi (t)\right\rangle ,%
\end{array}
\label{fidelity}
\end{equation}%
to clarify our dynamical method. We plot these quantities for three typical
conditions in Fig. \ref{fig6}. Numerical results indicate that a pair of
electrons can be near perfectly oscillates between two unit cells with
different periods. Here the numerical simulation is only performed for the
case with $n=0$, which corresponds to $U_{\text{eff}}/J_{\text{eff}}=2.309$
from Eq. (\ref{ratio}). This ratio is much larger than $U/\kappa =0.010\sim
0.023$ indicated in Fig. \ref{fig6}. Numerical results illustrate that a
spatially periodic external field really plays an important role in
enhancing the correlation effect. The setting of system parameters is
expected to provide guidance for the experiment.

\section{Summary}

\label{Summary}

In summary, we demonstrated that the band structure of a simple
one-dimensional Hubbard model can be constructed by applying a commensurate
external field. Such a moir\'{e} Hubbard model has a midgap flat band for
zero $U$. The midgap physics can be well described by an effective Hubbard
model when a weak $U$ is switched on. The effective strengths of hopping and
Hubbard interaction are controlled by the amplitude and period of the
external field. This makes it possible to observe strongly correlated
behaviors in a weak correlated system by adding proper spatially%
periodic external field. Numerical simulations are performed for the
finite-sized moir\'{e} Hubbard system to demonstrate the validity of this
approach. (i) We demonstrated that there exists a set of bound-pair energy
levels, including the $\eta $-pairing state. It indicates that the
tight-binding approximation works well. (ii) We further presented the
phenomenon of bound-pair oscillation, which reveals the dynamics occuring in
a moir\'{e} Hubbard system with relative strong correlation. The%
phenomenon of oscillation indicates that the on-site Hubbard interaction in
the unit of hopping strength can be enhanced by more than $2$ orders. This
work provides a method to enhance the correlated effect by the spatially
periodic external field and is expected as guidance for the experiment.

\acknowledgments This work was supported by National Natural Science
Foundation of China (under Grant No. 11874225).

\end{document}